\documentclass[%
preprint,
showpacs,
showkeys,
preprintnumbers,
nofootinbib,
amsmath,amssymb,amsfonts,
aps,
prd,
floatfix,a4paper,11pt]{revtex4-1}
\usepackage{amsmath}
\usepackage{amssymb}
\usepackage{color}
\usepackage{CJK}
\usepackage{graphicx}
\usepackage{dcolumn}
\usepackage{bm}
\usepackage[dvipdfm,bookmarks=true,colorlinks,%
            citecolor=blue,linkcolor=blue, hypertex, %
            breaklinks=true]{hyperref}

\begin{document}
\preprint{MPP-2012-82}
\title{Holographic superconductors with $z=2$ Lifshitz scaling}
\date{\today}
\author{Yanyan Bu}
\email{yybu@mpp.mpg.de}
\affiliation{State Key Laboratory of Theoretical Physics, Institute of
Theoretical Physics, Chinese Academy of Science, Beijing 100190, People's
Republic of China}
\affiliation{Max-Planck-Institut f$\ddot{u}$r Physik
(Werner-Heisenberg-Institut),
F$\ddot{o}$hringer Ring 6, 80805 M$\ddot{u}$nchen, Germany}
\begin{abstract}
We use gauge/gravity duality to explore strongly coupled superconductors with
dynamical exponent $z=2$. In the probe limit we numerically establish background
solutions for the matter fields and plot the condensate versus the dimensionless
temperature. We then investigate electromagnetic perturbations in order to
compute the AC conductivity and also calculate the spectral function. Our
results for the condensate and conductivity are qualitatively similar to those
of the AdS superconductor. However, we find that (for both s- and p-wave) the
condensate does not approach a constant at very low temperature and the
conductivity goes to one from below but never exceeds it in the high frequency
limit, in contrast to the AdS black hole. We do not see a peak at nonzero
frequency in the imaginary part of the AC conductivity along the $x$ direction
for the p-wave case. These features are due to the nontrivial dynamical
exponent. To be specific, the black hole geometry considered in this work is
anisotropic between space and time, very different from the Schwarzschild-AdS
black hole, which results in different asymptotic behaviors of temporal and
spatial components of gauge fields than those in the Schwarzschild-AdS black
hole.
\end{abstract}
\pacs{11.25.Tq}

\maketitle

\tableofcontents

\section{Introduction} \label{section1}
The application of gauge/gravity duality \cite{hep-th/9711200} to the
investigations of strongly coupled systems has gained broad interest ranging
from QCD phenomena at low energy to strongly correlated condensed matter
physics, see, e.g.,
\cite{0711.4467,*0901.0935,*0903.3246,*0904.1975,*0904.2750,*0909.0518} for some
recent reviews. Holographic superconductors have been constructed in
\cite{0803.3295,*0810.1563,0805.2960} by putting the Abelian Higgs model or
SU(2) gauge field into the AdS black hole spacetime. When the Hawking
temperature is decreased to a critical value, the black hole becomes unstable
against small perturbations and develops hair by condensing some field to
stabilize the system. This can be considered as the holographic realization of
the superconducting phase transition. This kind of construction for holographic
superconductors takes the (asymptotically) AdS black hole spacetime as the
starting point. According to the AdS/CFT correspondence, the AdS black hole
geometry
corresponds to a relativistic CFT at finite temperature. However, many condensed
matter systems do not have relativistic symmetry and it is therefore very
natural and interesting to generalize these holographic superconducting models
to non-relativistic situations.

On the other hand, inspired by the dynamical exponent in condensed matter
physics near the critical point, many papers have appeared on the
construction of black hole geometries with anisotropic scaling, such as the
Lifshitz black hole. The dual geometry of Lifshitz fixed points was first
proposed in \cite{0808.1725} and then generalized to finite temperature in
\cite{0905.3183,*0812.0530}. In short, we now have a geometrical realization of
a strongly coupled anisotropic field theory at finite temperature. It is
expected that non-relativistic AdS/CFT will help us to understand some puzzles
in unconventional condensed matter physics\footnote{Actually, there appear many
works by taking this kind of non-relativistic metrics to reveal some strange
features of condensed matter system, see \cite{0809.2020,*0912.2403,*1003.5361,
*1007.0590,*1010.6075,*1105.3197,*1106.2291,*1112.0573,*1203.1367} for an
incomplete list.}.

In \cite{0912.1061} a variety of strange metallic behaviors have been realized
by using gravity duals of the Lifshitz fixed points and some string theoretical
realizations of this geometry have been proposed in that paper\footnote{Another
string theory realization of Lifshitz-like fixed points was investigated in
\cite{0905.0688}.}. More recently, holographic fermions have been studied in
\cite{1201.3832,*1201.1764} to produce a non-Fermi liquid behavior. With these
studies in mind, we now generalize holographic superconductors under the
relativistic AdS/CFT framework to the Lifshitz black hole geometry in order to
explore the effects of the dynamical exponent and also in the hope to
distinguish some universal properties of holographic superconductors. In actual
fact, holographic s-wave superconductors with a Lifshitz fixed point have been
constructed in \cite{0909.4857,0908.2611} and in
Ho$\check{\text{r}}$ava-Lifshitz gravity in \cite{0911.4867}. However, the work
of \cite{0908.2611} produced only the condensate and the results of
\cite{0909.4857} seemed surprising, especially the very small real part of the
AC conductivity. We therefore more carefully study these systems here in order
to clarify some confusions which appeared in \cite{0909.4857} and also reveal
some properties of p-wave superconductors with Lifshitz scaling.

Although there is a dynamical exponent, which makes the Lifshitz geometry behave
quite differently from asymptotically AdS spacetime, we find that qualitative
behaviors of holographic superconductors with Lifshitz scaling (for both s-wave
and p-wave) are basically the same as those of the AdS black hole case. The
condensate has mean-field behavior near the critical temperature. A gap does
appear when decreasing the temperature, which is manifest once looking at the
real part of the AC conductivity. There is also a delta peak near zero frequency
for the conductivity, which is a signal of DC superconductivity. The real part
of the conductivity approaches one in the high frequency limit where the
imaginary part goes to zero. These common characteristics of holographic
superconductors appear to be robust phenomena and can be taken as universal
properties of gauge/gravity duality when applied to the study of strongly
coupled condensed matter systems.

We also find some other interesting features in our work. The first is that all
the condensates do not approach some constant in the zero temperature limit
compared to the BCS superconductor and the AdS black hole holographic
superconductors. The real part of the conductivity never exceeds one and the
imaginary part of the conductivity just approaches zero from above but never
goes below zero in the high frequency limit. What is more striking is that there
is no pole at nonzero frequency for the imaginary part of the AC conductivity
$\sigma_{xx}(\omega)$ in the p-wave case. We therefore attribute these
nontrivial features to the effect of the Lifshitz scaling. More specifically,
the black hole geometry considered in this work is anisotropic between space and
time, very different from the AdS black hole, which results in different
asymptotic behaviors of temporal and spatial components of gauge fields than
previous conclusions in AdS black hole.

Our paper is organized as follows. In section~\ref{section2} we briefly review
basic aspects of asymptotically Lifshitz black holes for further studies.
Section~\ref{section3} is concerned with the construction of s-wave
superconductors in the Lifshitz black hole geometry. With the numerical
solutions one can plot the condensate as well as the free energy difference
between the normal and superconducting phases versus the dimensionless
temperature. We find that the free energy difference is always greater than zero
when below the critical temperature $T_c$, which proves that the superconducting
phase is at least thermodynamically stable. Then we move on to investigate the
electromagnetic fluctuations of the system and numerically calculate the AC
conductivity using linear response theory. The spectral functions for the
electromagnetic perturbations are calculated as well. In section~\ref{section4}
the corresponding results for the p-wave superconductor are presented. A short
summary is given in section~\ref{section5}.

\section{Holographic setup: the gravity dual of the Lifshtiz fixed
point}\label{section2}
In this section, we concisely provide some backgrounds for the gravity dual of
the Lifshitz fixed point. As mentioned in section~\ref{section1}, there exist
field theories with anisotropic scaling symmetry between the temporal and
spatial coordinates. This is found for example in some condensed matter systems
near the critical point,
\begin{equation}
t\rightarrow \lambda^z t,\quad x^i\rightarrow \lambda x^i,
\end{equation}
where $z$ is called the dynamical exponent. One geometrical realization of this
scaling symmetry comes from the generalized gauge/gravity duality: we can map
this scaling symmetry in the field theoretical side to some geometrical symmetry
in the gravity side. Then it is straightforward to write down the metric with
this type of scaling symmetry,
\begin{equation}
ds^2=L^2\left(-r^{2z}dt^2+r^2\sum_{i=1}^{d}dx_i^2+\frac{dr^2}{r^2}\right),
\end{equation}
where $0<r<\infty$ and $L$ is the radius of curvature of the geometry. This
geometry was first proposed in \cite{0808.1725} where the action sourcing this
geometry was also given. The scale transformation takes the following form,
\begin{equation}
t\rightarrow \lambda^z t,\quad x^i\rightarrow \lambda x^i,\quad r\rightarrow
\frac{r}{\lambda}.
\end{equation}
When $z=1$, the above geometry reduces to the usual $AdS_{d+2}$ spacetime and
the symmetry group is enlarged to $SO(d+1,2)$.

According to the gauge/gravity duality, to put the dual field theory at finite
temperature one can study the metric with a black hole and the Hawking
temperature of the black hole is identified as the temperature of the dual field
theory. It is therefore of great interest to construct a finite temperature
version of the above geometry, i.e. the so-called Lifshitz black hole. However,
it is difficult to obtain analytic black hole solutions in Lifshitz spacetimes.
There are some attempts to construct Lifshitz black holes\footnote{See
\cite{0812.5088,*0905.1136,*0905.2678,*0908.1272,*0909.0263,
*0909.1347,*0909.2807,*0910.4428,*0911.2777,*0911.3586,*1001.2361,*1001.4945,
*1002.4448,
1003.5064,*1005.3291,*1007.2490,*1008.2062,*1009.3445,*1102.0578,*1102.5344,
*1105.4862,
*1105.6335,*1107.4451,*1201.1905,*1202.1748,*1203.0576} for an incomplete list
of the constructions of black hole geometries with anisotropic scaling
symmetry.} and we here follow the work of \cite{0812.0530}. It was found that
action of the form
\begin{equation}
S=\frac{1}{16 \pi G_{d+2}}\int
d^{d+2}x\sqrt{-g}\left(R-2\Lambda-\frac{1}{2}\partial_{\mu}\phi\partial^{\mu}
\phi- \frac{1}{4}e^{\lambda\phi}\mathcal {F}_{\mu\nu}\mathcal
{F}^{\mu\nu}\right),
\end{equation}
where $\Lambda$ is the cosmological constant, $\phi$ is a massless scalar and
$\mathcal{F}_{\mu\nu}$ is an abelian gauge field strength, admits the following
black hole geometry,
\begin{equation}
ds^2=L^2\left(-r^{2z}f(r)dt^2+r^2\sum_{i=1}^{d}dx_i^2+\frac{dr^2}{r^2f(r)}
\right), \end{equation}
\begin{equation}
f(r)=1-\frac{r_0^{z+d}}{r^{z+d}},\quad \Lambda=-\frac{(z+d-1)(z+d)}{2L^2}.
\end{equation}
To support the above black hole geometry, one also needs to give the backgrounds
for $\phi$ and $\mathcal{F}_{rt}$,
\begin{eqnarray}
&e^{\lambda \phi}=r^{-2d},\quad \lambda^2=\frac{2d}{z-1},\nonumber\\
&\mathcal{F}_{rt}=q_0 r^{z+d-1},\quad q_0^2=2L^2 (z-1)(z+d).
\end{eqnarray}

Evidently, choosing the dynamical exponent $z$ to be 1 reduces the Lifshitz
black hole to the Schwarzschild AdS black hole in $d+2$-dimensions. The Hawking
temperature and the entropy of the black hole are,
\begin{equation}
T=\frac{z+d}{4\pi}r_0^z,\quad S_{en}=\frac{L^d V_d}{4G_{d+2}}r_0^d,
\end{equation}
where $V_d$ denotes the volume of the d-dimensional space.

We do a coordinate transformation $u=r_0/r$ to map the holographic direction $r$
into a finite interval [0,1] as we find that it is more convenient to use this
coordinate system when carrying out numerical calculations. What's more, we will
focus on a 4-dimensional bulk theory and choose the dynamical exponent $z=2$.
With these choices, the bulk geometry is reduced to,
\begin{equation}
ds^2=L^2\left(-\frac{r_0^{2z}}{u^{2z}}f(u)dt^2+\frac{r_0^2}{u^2}(dx^2+dy^2)+
\frac{du^2}{u^2f(u)}\right),\quad f(u)=1-u^{z+2}\label{Lifshitz bk}.
\end{equation}
In this new coordinate system, the horizon is located at $u=1$ and $u=0$ denotes
the conformal boundary where the dual field theory lives.
\section{Holographic s-wave superconductors with Lifshitz
scaling}\label{section3}
This section and the next are the central parts of our work. In this section we
focus on aspects of s-wave superconductor. In subsection~\ref{subsection1} we
list the equations of motion for the background fields and solve them by a
shooting method. Subsection~\ref{subsection2} is devoted to the studies of the
electromagnetic perturbations of the system.
\subsection{Solution for the background fields}\label{subsection1}
In \cite{0801.2977}, it was shown that a charged AdS black hole supports charged
scalar hair if the charge is large enough. Later this idea was generalized to
the neutral AdS black hole in \cite{0803.3295,*0810.1563}, which is the first
model of an s-wave holographic superconductor. More specifically, it is
constructed
from the Abelian Higgs model in the AdS black hole background. The action for
this system is,
\begin{equation}
S=\int
d^4x\sqrt{-g}\left(R+\frac{6}{L^2}-\frac{1}{4}F_{\mu\nu}F^{\mu\nu}
-\left\vert\partial_{\mu}
\Psi-iqA_{\mu}\Psi\right\vert^2-V(\left\vert\Psi\right\vert)\right),\label{
action swave}
\end{equation}
where we choose $V(\left\vert\Psi\right\vert)=m^2\left\vert\Psi\right\vert^2$
for simplicity. This is the minimal Lagrangian of the gravitational dual which
holographically describes a superconducting phase transition. As mentioned
before, we work in the probe limit, i.e, the black hole geometry is fixed and
feels no effect of the matter fields. Above the critical temperature, the black
hole background is stable and the scalar field $\Psi$ can be set to zero. This
corresponds to the normal phase. When the temperature is decreased to the
critical value, the black hole background becomes unstable against small
perturbations and the scalar field wants to condense in order to stabilize the
system. Once this happens, the black hole develops hair and the system goes
through a superconducting phase transition. The scalar field $\Psi$
holographically models the order parameter of a conventional superconductor and
gets a nontrivial profile only in the superconducting phase. As the AdS/CFT
correspondence maps a strongly coupled field theory to a weakly coupled gravity
system, the holographic method is expected to give a description of strongly
coupled superconductors in contrast to conventional BCS theory. However, it is
still far from being clear as to the pairing mechanism in holographic
superconductors. Due to the conformal characteristic of the AdS space, the
chemical potential is usually introduced to explicitly break the conformal
invariance and to make the temperature scale meaningful. This is achieved by
turning
on the time component of the U(1) gauge field.

Having fixed the black hole geometry, the equations of motion for $A_{\mu}$ and
$\Psi$ can be easily extracted from the Euler-Lagrange equations and are listed
here for further studies in later subsections,
\begin{eqnarray}
\frac{1}{\sqrt{-g}}\partial_{\mu}\left(\sqrt{-g}F^{\mu\nu}\right)=iq[
\Psi^*(\partial^{\nu} \Psi-iq
A^{\nu}\Psi)-\Psi(\partial^{\nu}\Psi^*+iqA^{\nu}\Psi^*)],\label{eq for at}\\
\partial_{\mu}[\sqrt{-g}(\partial^{\mu}\Psi-iqA^{\mu}\Psi)]=\sqrt{-g}[V^{\prime}
(\left\vert \Psi\right\vert)\frac{\Psi}{2\left\vert\Psi\right\vert}+iq
A^{\mu}(\partial_{\mu}\Psi-iqA_{\mu}\Psi)].
\end{eqnarray}
From the radial component of eq.~(\ref{eq for at}) one can show that the scalar
field can be taken as real. Therefore, the ansatz for the backgrounds of the
gauge and scalar fields are
\begin{equation}
A=\phi(u)dt,\quad \Psi=\psi(u).
\end{equation}
Recalling the Lifshitz black hole background given in eq.~(\ref{Lifshitz bk}),
we obtain the following equations of motion,
\begin{equation}
\phi^{\prime\prime}+\frac{z-1}{u}\phi^{\prime}-\frac{2\psi^2}{u^2f(u)}\phi=0,
\label{eqa0}
\end{equation}
\begin{equation}
\psi^{\prime\prime}+\left[\frac{f^{\prime}(u)}{f(u)}-\frac{z+1}{u}\right]\psi^{
\prime}+\frac{u^{2z-2}
\phi^2}{f^2(u)}\psi-\frac{m^2L^2}{u^2f(u)}\psi=0,\label{eqa1}
\end{equation}
where we have for simplicity chosen the scalar potential $V(\Psi)$ as
$m^2\Psi^2$, giving mass to $\Psi$, and set the charge $q^2L^2=1$. The prime in
these equations denotes derivative with respect to $u$ and this notation will be
used in the following.

Before continuing, some remarks are in order. Actually, in
eqs.~(\ref{eqa0},\ref{eqa1}), we have rescaled the fields $\phi$ and $\psi$ for
convenience. When the dynamical exponent $z$ is nontrivial, the rescaling is
different between $\psi$ and $\phi$,
\begin{equation}
\phi\rightarrow \frac{z+2}{4\pi T}\phi,\quad \psi\rightarrow
\left(\frac{z+2}{4\pi T}\right)^{1/z}\psi.
\end{equation}
The superconducting phase transition is a spontaneous breaking of the
electromagnetic symmetry, which should be reflected in the holographic
description in some sense. Strictly speaking, the gauge symmetry in the bulk
corresponds to a global symmetry on the boundary filed theory according to
gauge/gravity duality and therefore we have no electromagnetic symmetry on the
boundary. Then the holographic superconductor model should at most be thought of
as holographic superfluidity. However, as discussed in
\cite{0803.3295,*0810.1563} this model can produce many features of
superconductors and we can ignore this subtlety.

We now have a look at the asymptotic behaviors of the background fields $\phi$
and $\psi$. Near the horizon, one must have $\phi(1)=0$ for its norm to be
finite and the scalar field should also be finite there. Near the conformal
boundary, we have the following asymptotic behavior from the Frobenius analysis
of eqs.(\ref{eqa0}) and (\ref{eqa1}) near the singularity $u=0$,
\begin{equation}
\phi(u\rightarrow 0)\sim \mu+\rho u^{2-z} \quad (z\neq 2) \quad\text{or}\quad
\phi(u\rightarrow 0)\sim \rho+\mu \log u \quad (z =2),
\end{equation}
and
\begin{equation}
\psi(u\rightarrow 0)\sim \psi^{(\nu_+)}u^{\nu_+}+\psi^{(\nu_-)}u^{\nu_-},
\end{equation}
where the scaling dimension $\nu_{\pm}$ of the scalar operator $\mathcal{O}$
dual to the bulk scalar $\psi$ is given by
$\nu_{\pm}=\frac{z+2\pm\sqrt{(z+2)^2+4m^2}}{2}$. It is explicit that the BF
bound for the scalar mass in Lifshitz background is now changed to $m^2 \geq
-\frac{(z+2)^2}{4}$ for 4-dimensional bulk. To obtain explicit behavior for
these two fields near the conformal boundary, we need to specify the mass square
of the scalar field and the dynamical exponent. We here choose $z=2$ and
$m^2=0,-3$ (which are both above the BF bound).

For the exponent $z=2$, the boundary behavior of the $\phi(u)$ field has a
logarithmic term\footnote{This kind of asymptotic behavior for the bulk field also happens in the study of p-wave holographic superconductor by taking the D3/D3 model in \cite{1205.1614}.} due to degenerate indices of eq.~(\ref{eqa0}) near the
conformal boundary $u=0$. Now the physical result is also changed: we should
identify the constant term $\rho$ as the charge density as it is a normalizable
mode with respect to the logarithmic term and the non-normalizable logarithmic
term $\mu$ as the chemical potential. Fortunately, this happens only for the
time component of the gauge field and the final result for the conductivity does
not change too much. We will come to see this when studying the AC conductivity
along the $x$ direction for the p-wave case.

From the boundary behavior of the scalar field $\psi$, we can directly read off
the expectation value of the dual operator $\mathcal{O}$. When $m^2=0$, the
indices $\nu=0$ and $4$,
\begin{equation}
\psi(u\rightarrow 0)\sim \psi^{(0)}+\psi^{(4)}u^4.
\end{equation}
To make the superconducting phase transition a spontaneous breaking of symmetry,
we should impose that
\begin{equation}
\psi^{(0)}=0 \quad \text{and} \quad \langle\mathcal{O}_4\rangle \sim \psi^{(4)}.
\end{equation}
Therefore, there is only one theory for this choice of the scalar mass squared.

While for $m^2=-3$ the indices $\nu=1$ or $3$,
\begin{equation}
\psi(u\rightarrow 0)\sim \psi^{(1)}u +\psi^{(3)}u^3.
\end{equation}
Evidently, the two modes in the above equation are both normalizable according
to \cite{hep-th/9905104}, and in order to make the theory stable, we should
either impose
\begin{equation}
\psi^{(1)}=0 \quad \text{and} \quad \langle\mathcal{O}_3\rangle \sim \psi^{(3)}
\end{equation}
or
\begin{equation}
\psi^{(3)}=0 \quad \text{and} \quad \langle\mathcal{O}_1\rangle \sim \psi^{(1)}.
\end{equation}
We now have two theories for $m^2=-3$ corresponding to dimension 1 or dimension
3 order parameters, respectively. In the following analysis, we will concentrate
on the dimension 3 case because we find that the ground state of the dimension 1
theory is numerically unstable.

\begin{figure}[h]
\includegraphics[scale=1.2]{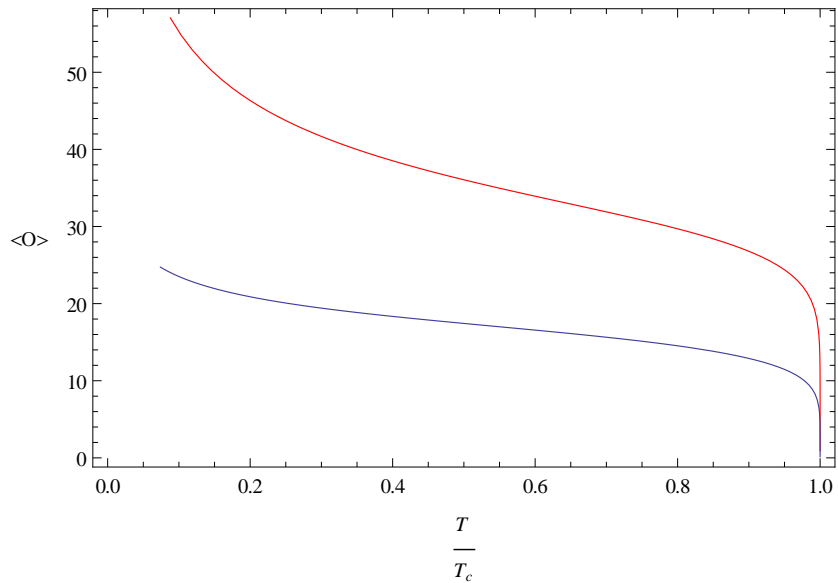}
\caption{The condensates of the s-wave superconductor for the dimensionless 3D
operator
$\mathcal{O}_3$ (blue): $\sqrt[3]{\langle\mathcal{O}_3\rangle}/T_c$ and 4D
operator
$\mathcal{O}_4$ (red): $\sqrt[4]{\langle\mathcal{O}_4\rangle}/T_c$.}
\label{figure1}
\end{figure}

With the boundary conditions mentioned above, we can now use a numerical
shooting method to solve the coupled nonlinear eqs.~(\ref{eqa0},\ref{eqa1}). The
condensates corresponding to operators $\mathcal{O}_3$ and $\mathcal{O}_4$ are
plotted in FIG.~\ref{figure1}.

From this figure, we see that the condensates go to zero at the critical
temperature $T_c$. However, they do not approach some fixed constants as the
temperature $T\rightarrow 0$, which is different from both conventional BCS
theory of weakly coupled superconductors and AdS black hole holographic
superconductors. We may attribute this effect to the nontrivial dynamical
exponent $z\neq1$. We also find that the expectation values for the operators
$\sqrt[3]{\langle\mathcal{O}_3\rangle}$ and
$\sqrt[4]{\langle\mathcal{O}_4\rangle}$ are much larger than the BCS predictions
at zero temperature. This is consistent with the results of AdS black hole
holographic superconductors. Perhaps this is due to the fact that the strongly
interacting nature of the holographic superconductor in contrast to the BCS
theory.

In the mean field theory for the superconductor, the order parameters have a
square root behavior near the critical temperature $T_c$,
\begin{equation}
\langle\mathcal{O}\rangle\sim (T_c-T)^{1/2}.
\end{equation}
By fitting these curves, we find the mean field behavior also holds in our
results. Specifically, for the dimension 3 theory:
\begin{equation}
\langle\mathcal{O}_3\rangle\approx (18.9117T_c)^3(1-T/T_c)^{1/2} \quad
\text{when} \quad T\rightarrow T_c,
\end{equation}
where the critical temperature $T_c\approx 0.0351935 \mu$. For the dimension 4
theory:
\begin{equation}
\langle\mathcal{O}_4\rangle\approx (35.5940T_c)^4(1-T/T_c)^{1/2} \quad \text{as}
\quad T\rightarrow T_c,
\end{equation}
with the critical temperature $T_c\approx0.0229931\mu$.

\begin{figure}[h]
\includegraphics[scale=1.2]{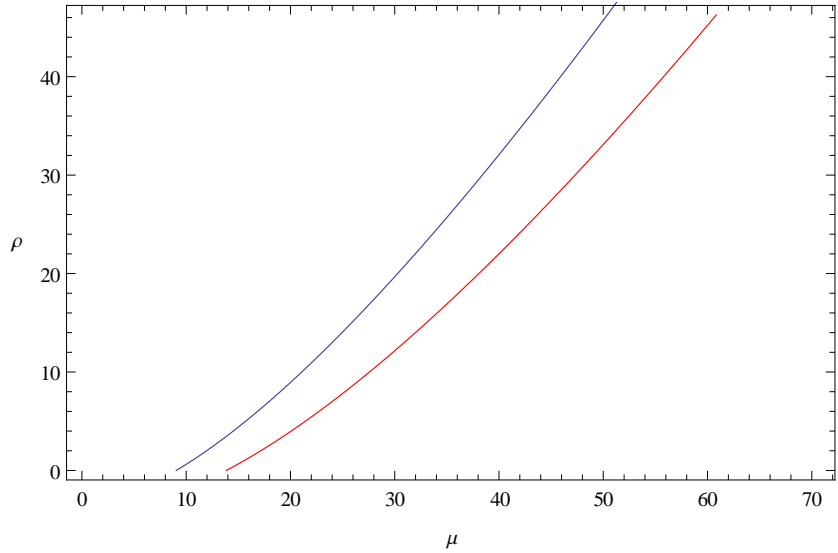}
\caption{The profile of the charge density as a function of the chemical
potential in the superconducting phase for the s-wave superconductors: the blue
one is for the dimension 3 theory and the red one is for the dimension 4
theory.}
\label{figure2}
\end{figure}

The relation between the chemical potential and charge density is plotted in
FIG.~\ref{figure2}. We see that the profile has a small deviation from the
linear behavior in the superconducting phase. There is one critical value for
the chemical potential where the charge density becomes nonzero. This is
actually the critical point where the superconducting phase transition occurs.

\begin{figure}[h]
\includegraphics[scale=1.2]{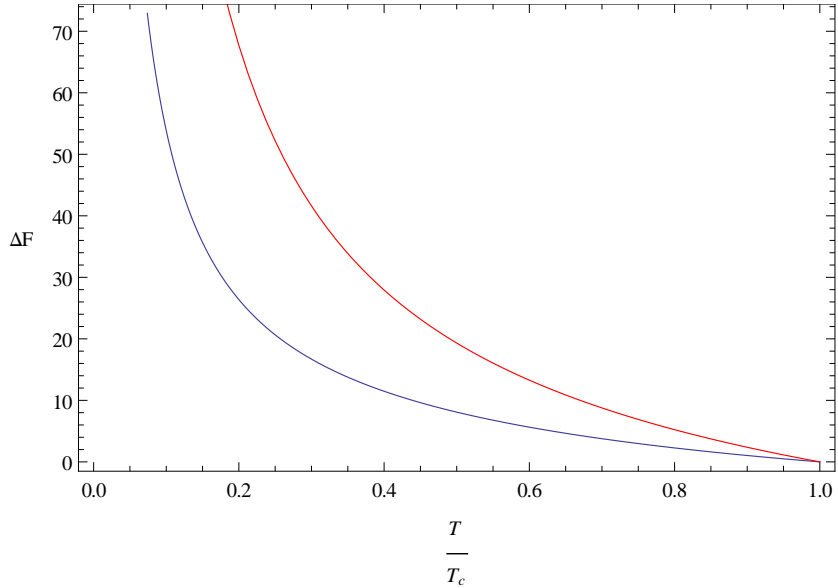}
\caption{The dimensionless free energy difference between the normal and the
superconducting phases of the s-wave superconductor for $\mathcal{O}_3$ (blue)
and $\mathcal{O}_4$ (red) theories, respectively.}
\label{figure3}
\end{figure}

At the end of this subsection, we plot the free energy for the two theories,
which can be taken as one piece of evidence that the superconducting phase
transition does happen when the temperature is decreased to the critical one.
With the equations of motion for the backgrounds $\phi$ and $\psi$, we can
reduce the action (\ref{action swave}) to some simpler expression by integrating
by parts. Then, the free energy difference between the normal and
superconducting phases is
\begin{equation}
\Delta \Xi_{N-SC}=\frac{V_2}{2g_4^2}r_0^{z+2}\int _0^1 du
\frac{u^{z-1}}{f(u)}(\phi(u)\psi(u))^2\equiv\frac{V_2}{2g_4^2}r_0^{z+2} \Delta
F.
\end{equation}
FIG.~\ref{figure3} is for the plot of the free energy difference.
Clearly, the superconducting phase is thermally stable when $T< T_c$.
\subsection{Fluctuation analysis: AC conductivity and the spectral
function}\label{subsection2}
To calculate the AC conductivity of this system, we need to study the
electromagnetic perturbation of the above background (here, we do not consider
fluctuation of the scalar field). For the s-wave superconductor, the
conductivity is isotropic and it is therefore equivalent to study any spatial
component of the gauge field. We here choose radial gauge for the
electromagnetic fluctuation, i.e., $a_u=0$. For the purpose of the AC
conductivity, we take the following ansatz for the electromagnetic perturbation
$a_x(t,\vec{x},u)$
\begin{equation}
a_x(t,\vec{x},u)=e^{-i\omega t}a_x(u),
\end{equation}
where we have taken spatial momentum to zero. Then, the equation of motion for
$a_x(u)$ is,
\begin{equation}
a_x^{\prime\prime}+\left[\frac{f^{\prime}(u)}{f(u)}-\frac{z-1}{u}\right]a_x^{
\prime}+
\left[\frac{\tilde{
\omega}^2u^{2z-2}}{f(u)^2}-\frac{2\psi^2}{u^2f(u)^2}\right]a_x=0,
\label{eq for ax}
\end{equation}
where the dimensionless frequency $\tilde{\omega}$ is defined as
$\tilde{\omega}=(z+2)\omega/4\pi T$. According to linear response theory, the
conductivity is given by the Kubo formula,
\begin{equation}
\sigma(\omega)=\frac{G(\omega,\vec{k}=0)}{i \omega},\label{kubo}
\end{equation}
where the retarded Green's function $G(\omega,\vec{k}=0)$ for the operator dual
to gauge field can be computed according to the prescription given in
\cite{hep-th/0205051}.

Near the horizon, we should take the ingoing wave boundary condition for the
electromagnetic field fluctuation in order to calculate the retarded Green's
function,
\begin{equation}
a_x(u)=(1-u)^{-i\tilde{\omega}/4}[1+a_x^1(1-u)+a_x^2(1-u)^2+a_x^3(1-u)^3+\cdots]
, \label{solution horizon}
\end{equation}
where we have set the scale of $a_x$ to be one by making use of the linearity of
the fluctuation equation. The coefficients in the above expansion can be
uniquely determined once the background $\psi$ and the frequency
$\tilde{\omega}$ are specified.

Near the conformal boundary $u=0$, the general solution to the fluctuation is of
the form,
\begin{equation}
a_x(u)=A_x^{0}+A_x^1u^2+\cdots.
\end{equation}
The conductivity can be expressed as
\begin{equation}
\sigma(\omega)=\frac{1}{i \omega}\frac{A_x^1}{A_x^0}
\end{equation}
by using the Kubo formula (\ref{kubo}). By the way, we also give the definition
of the spectral function $\mathcal{R}$ plotted later,
\begin{equation}
\mathcal{R}=-2\textrm{Im}G(\omega,\vec{k}=0) = -2 \frac{A_x^1}{A_x^0}.
\end{equation}

\begin{figure}[h]
\includegraphics[scale=0.8]{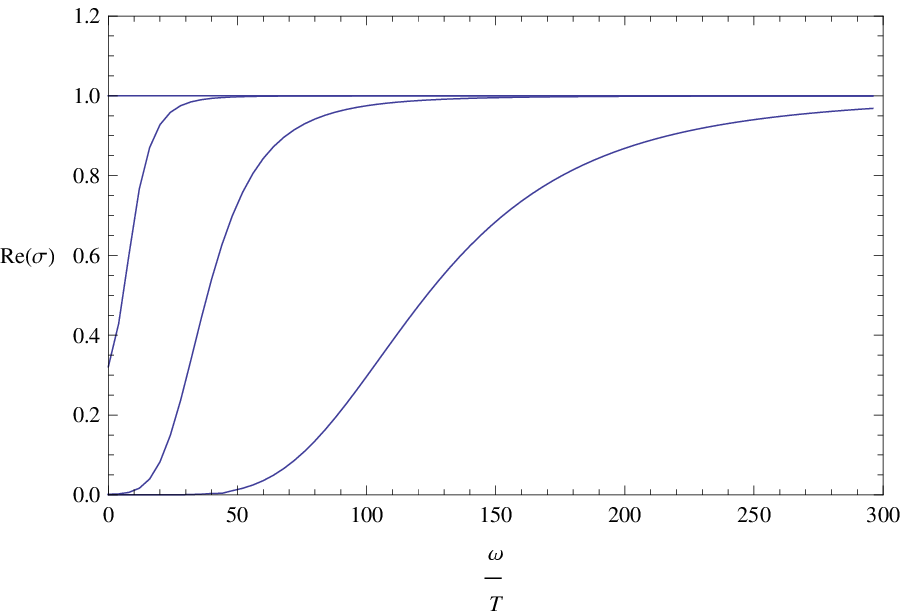}
\includegraphics[scale=0.8]{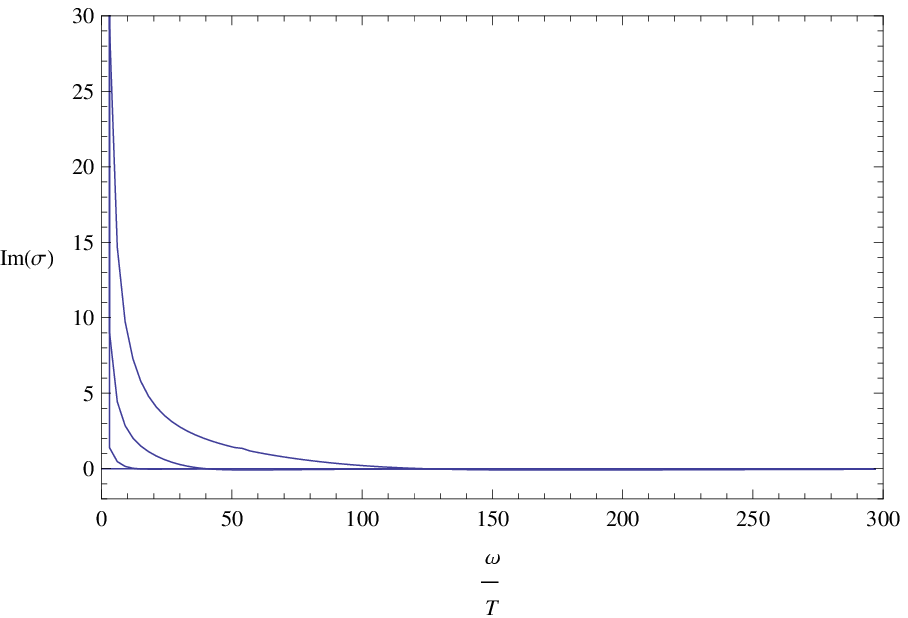}
\caption{The real (imaginary) part of the AC conductivity of the s-wave
superconductor for the dimension 3 theory at $T/T_c=1.0, 0.798694, 0.269798,
0.0864351$ from top to bottom (bottom to top).}
\label{figure4}
\end{figure}

\begin{figure}[h]
\includegraphics[scale=0.8]{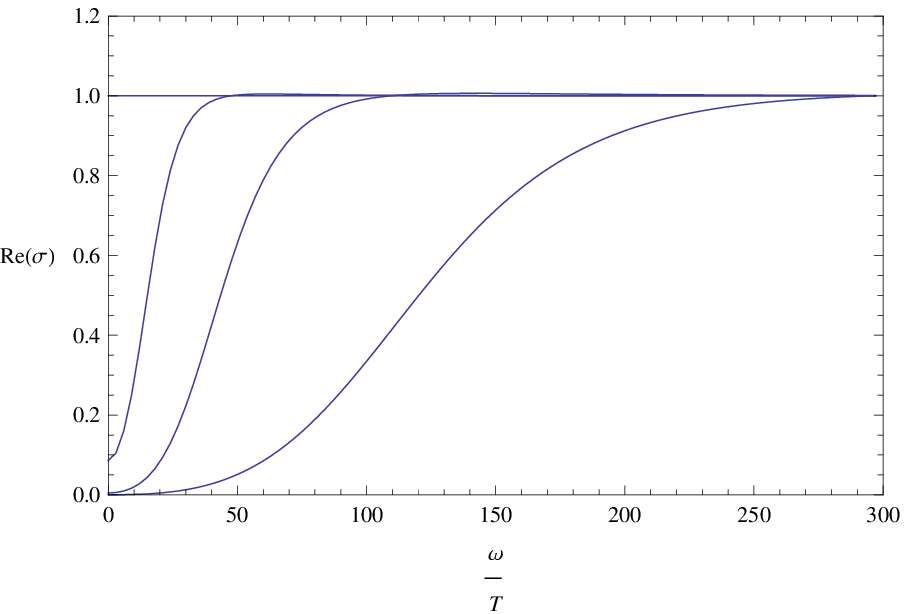}
\includegraphics[scale=0.8]{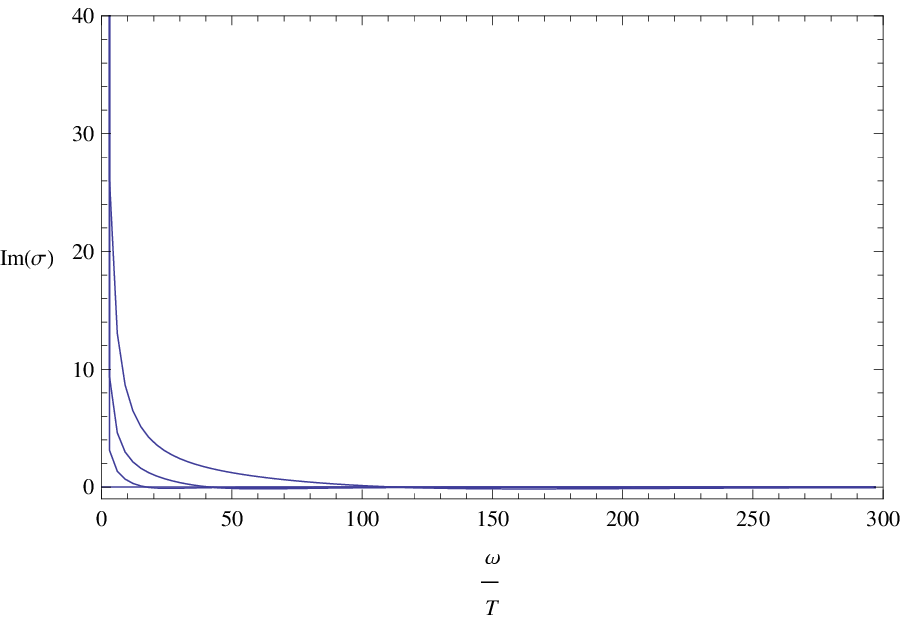}
\caption{The real (imaginary) part of the AC conductivity of the s-wave
superconductor for the dimension 4 theory at $T/T_c=1.0, 0.565265, 0.298191,
0.0879648$ from top to bottom (bottom to top).}
\label{figure5}
\end{figure}

Generically, the equation of motion for the electromagnetic fluctuation (\ref{eq
for ax}) cannot be solved analytically due to the presence of the scalar field.
The philosophy of the numerical method is that we can use the power series
solution as in eq.~(\ref{solution horizon}) and do numerical integration from
the
horizon to the conformal boundary. Then we can extract the coefficients $A_x^0$
and $A_x^1$ and therefore get the conductivity. Our numerical results for the AC
conductivity are plotted in FIG.~\ref{figure4} and FIG.~\ref{figure5}.
From the plots, we find that the real part of the conductivity approaches one in
the high frequency limit, which is equal to the results of the normal phase.
This is explicit when taking a look at eq.~(\ref{eq for ax}): the effect of the
condensed field $\psi$ can be neglected at large $\tilde{\omega}$. Compared to
the result for the conductivity in \cite{0909.4857}, our result is of order $1$
although these profiles are qualitatively similar. We found that the results for
dimension 3 and dimension 4 theory are qualitatively the same. We also see that
a gap forms as the temperature is lowered and the gap gets increasingly deep
until the conductivity is exponentially small, which is the same as the AdS
superconductor \cite{0803.3295,*0810.1563}.

In the imaginary part of the conductivity, there is a pole at zero frequency.
This can be explained from the Kramers-Kr\"{o}nig relation
\begin{equation}
\text{Im}[\sigma(\omega)]=-P\int_{-\infty}^{\infty}\frac{d\omega^{\prime}}{\pi}
\frac{
\text{Re}[\sigma(\omega^{\prime})]}{\omega^{\prime}-\omega},\label{kk relation}
\end{equation}
where $P$ denotes the principal value of the integration. From this formula, we
can see that the real part of the conductivity contains a delta function,
$\text{Re}[\sigma(\omega)]=\pi\delta(\omega)$, only when the imaginary part has
a pole, $\text{Im}[\sigma(\omega)] = 1/\omega$. Actually, there is a peak at
zero frequency for the imaginary part of the conductivity as can be seen in the
right plots of FIGs.~\ref{figure4} and ~\ref{figure5}.\footnote{In fact, we can
explicitly plot the delta peak in the real part of the conductivity. However,
when carrying out the numerical calculations, it is a little hard because we
find that this peak will appear at about $\omega\sim 10^{-20}$ and the numerical
computations are not stable there.}

\begin{figure}[h]
\includegraphics[scale=1.5]{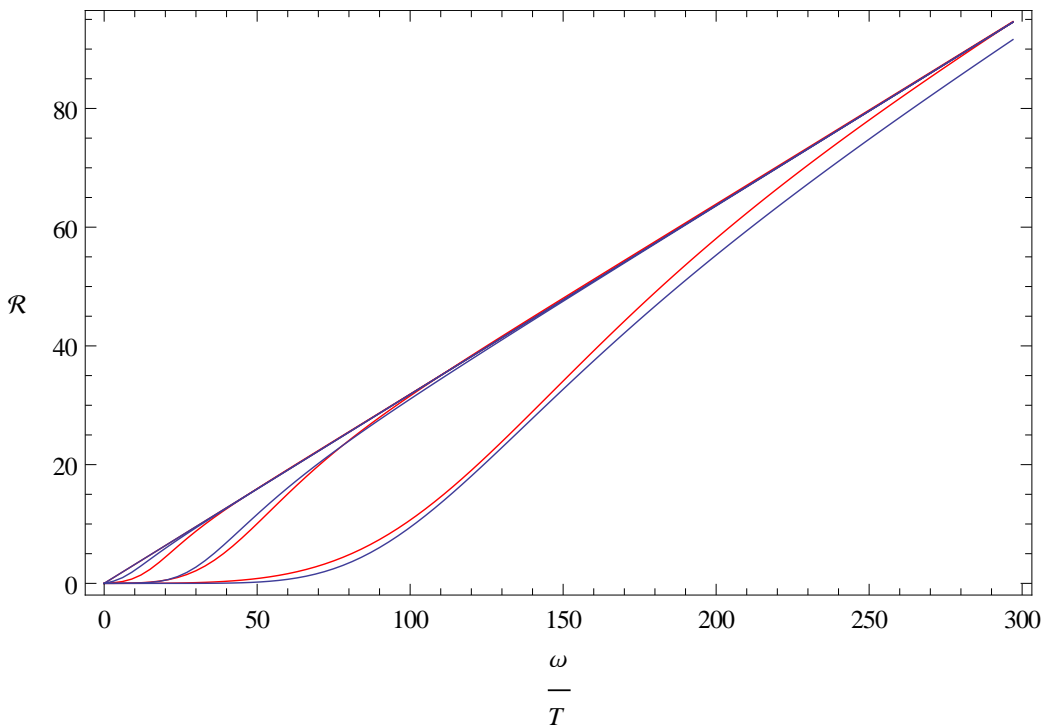}
\caption{The spectral functions of the s-wave superconductor for the dimension 3
theory (blue) at $T/T_c=1.0, 0.798694, 0.269798, 0.0864351$ and the dimension 4
theory (red) at $T/T_c=1.0, 0.565265, 0.298191, 0.0879648$ from top to bottom.}
\label{figure6}
\end{figure}

In FIG.~\ref{figure6} we plot the results for the spectral function
corresponding to the gauge field fluctuation $a_x$. The two figures are
consistent with the real part of the conductivity: a gap will appear when the
condensate is nonzero (equivalently, when the temperature is low enough) and the
spectral functions have a linear behavior with respect to the frequency at large
$\omega/T$ (which corresponds to large $\omega/T$ behavior of the conductivity).

\section{Holographic p-wave superconductors with Lifshitz
scaling}\label{section4}
In this section we present the results of the p-wave superconductor in the
Lifshitz black hole background. The structure of this section is the same as the
previous one.
\subsection{Solution for the background fields}
As we have seen in the last section, the s-wave holographic superconductor is
very simple and also elegant in describing some important features of
superconductors. However, the Abelian-Higgs model for the s-wave superconductor
appears to be less universal: we have to specify a potential for the scalar. On
the other hand, the p-wave superconductor has already been observed in some
experiments of condensed matter physics. It is natural as well as interesting to
extend the construction of holographic s-wave superconductors to the p-wave
situation. This has been achieved in \cite{0805.2960} by introducing the SU(2)
gauge field into the $AdS_4$ black hole background\footnote{Another approach to
p-wave superconductor where the boundary field theory is known is based on
D-brane probe in black brane geometry
\cite{0810.2316,*0903.1864,*0810.3970,*0907.1508}.}. In this p-wave model the
chemical potential and the order parameter have been unified into one field, the
nonabelian gauge field, and the action is uniquely determined by gauge
invariance. In this sense, the p-wave holographic superconductor is more
universal than the Maxwell-Scalar system for the s-wave superconductor. The
action is simply of the form,
\begin{equation}
S=\frac{1}{2\kappa^2}\int d^4x\sqrt{-g}[R+\frac{6}{L^2}-\frac{1}{4}F_{\mu\nu}^a
F^{a\mu\nu}],
\end{equation}
where the $SU(2)$ gauge field strength is defined as
$F_{\mu\nu}^a=\partial_{\mu}A_{\nu}^a-\partial_{\nu}A_{\mu}^a+\epsilon^{abc}A_{
\mu}^b A_{\nu}^c$. The Yang-Mills equation for the gauge field is
\begin{equation}
\partial_{\mu}(\sqrt{-g}F^{a\mu\nu})+\sqrt{-g}\epsilon^{abc}A_{\mu}^b
F^{c\mu\nu}=0. \label{Yang-Mills eq}
\end{equation}

The gauge field ansatz is taken as
\begin{equation}
A=\phi(u)\tau^3dt+\psi(u)\tau^1dx,
\end{equation}
where $\tau^a$ is the generator of the $SU(2)$ gauge group. Plugging this ansatz
into the Yang-Mills equation (\ref{Yang-Mills eq}) results in
\begin{equation}
\phi^{\prime\prime}+\frac{z-1}{u}\phi^{\prime}-\frac{\psi^2}{f(u)}\phi=0,\label{
eqa03}
\end{equation}
and
\begin{equation}
\psi^{\prime\prime}+\left[\frac{f^{\prime}(u)}{f(u)}+\frac{1-z}{u}\right]\psi^{
\prime}+\frac{u^{2z-2} \phi^2}{f(u)^2}\psi=0.\label{eqa11}
\end{equation}

\begin{figure}[h]
\includegraphics[scale=1.2]{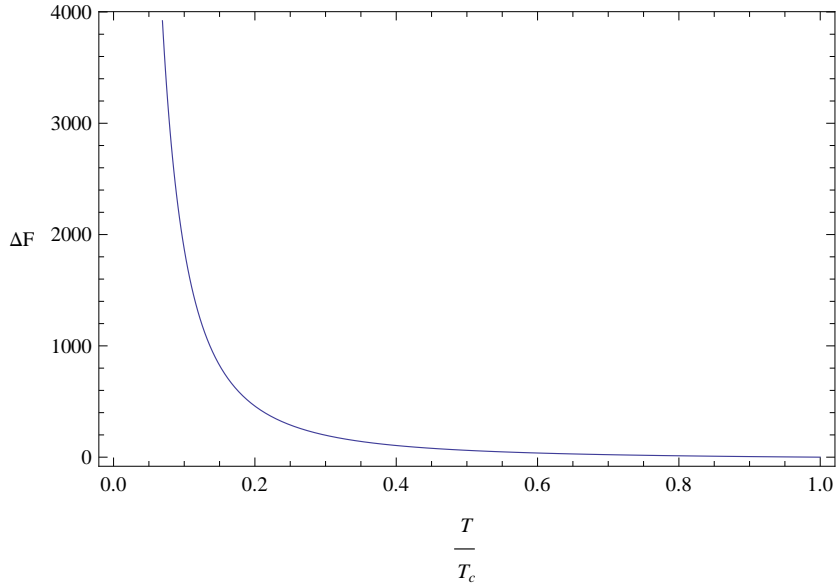}
\caption{The free energy difference between the normal and the superconducting
phases for the p-wave.}
\label{figure7}
\end{figure}

\begin{figure}[h]
\includegraphics[scale=1.1]{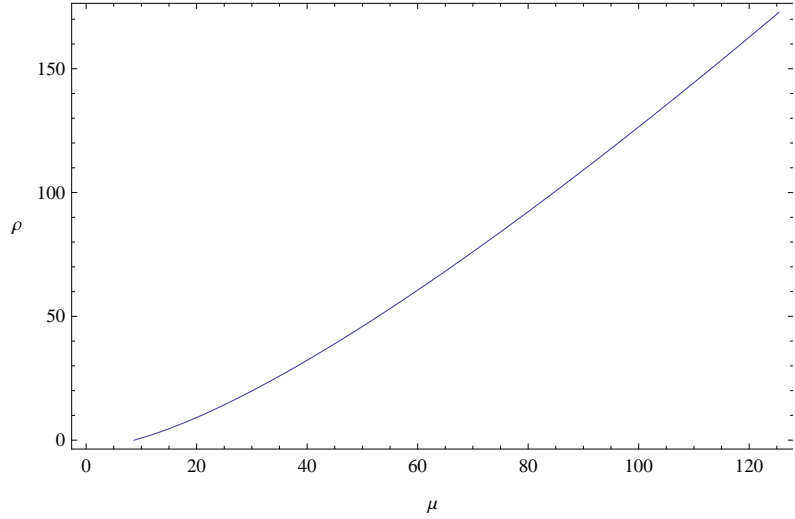}
\caption{Density as a function of $\mu$ for the p-wave superconductor.}
\label{figure8}
\end{figure}

As in the s-wave model, boundary behavior of the background fields $\phi(u)$ and
$\psi(u)$ is listed as below having chosen $z=2$,
\begin{eqnarray}
\phi(u\rightarrow 0)\sim \rho+ \mu \log u,\\
\psi(u\rightarrow 0)\sim \psi^{(0)}+\psi^{(2)}u^2.
\end{eqnarray}
The logarithmic term appears again for the $\phi$ field, which suggests that we
should identify the constant term $\rho$ as the charge density. We also impose
$\psi^{(0)}$ to be zero in our numerical calculations to insure that the
superconducting phase transition is a spontaneous breaking of the symmetry. Then
the value of $\psi^{(2)}$ is proportional to the expectation value of the p-wave
order parameter, i.e. $\langle\mathcal{O}_2\rangle\sim \psi^{(2)}$. Our
numerical results are plotted in the following figures. The free energy
difference
between the normal and superconducting phases is plotted in FIG.~\ref{figure7}.
As in the s-wave model, the superconducting phase is stable compared to the
normal phase when $T<T_c$. The charge density as a function of the chemical
potential can be found in FIG.~\ref{figure8}. Explicitly, these plots are very
similar to the s-wave case.

\begin{figure}[h]
\includegraphics[scale=1.2]{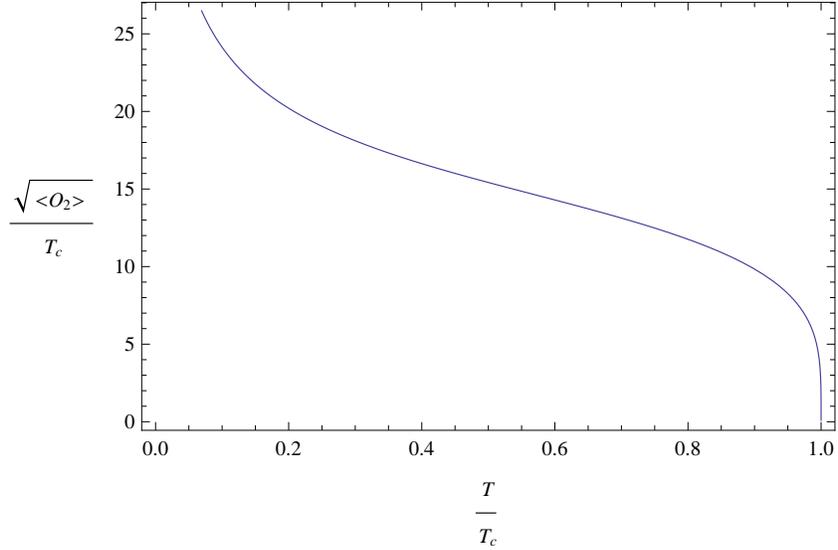}
\caption{The condensate $<\mathcal{O}_2>$ of the p-wave superconductor.}
\label{figure9}
\end{figure}

The condensate of the p-wave order parameter $\langle\mathcal{O}_2\rangle$ is
plotted in FIG.~\ref{figure9}. It is clear to see from this plot that the
condensate does not approach some fixed value when $T\rightarrow 0$, which is
the same as the results of the s-wave model mentioned in the last section. Once
again, this value is much larger than that of the AdS superconductor. By fitting
the curve near the phase transition, we find a mean field behavior for the
condensate,
\begin{equation}
\langle\mathcal{O}_2\rangle\approx (17.8217T_c)^2(1-T/T_c)^{1/2} \quad
\text{when}\quad T\rightarrow T_c,
\end{equation}
where the critical temperature is $T_c\approx 0.0367064\mu$.
\subsection{Fluctuation analysis: AC conductivities and spectral functions}
Owing to the appearance of the background for the gauge field, say $A_0^3$, the
SU(2) gauge symmetry is explicitly broken to the $U(1)_3$ generated by the
rotation in the colored $12$ plane. This residual symmetry is identified as the
electromagnetic symmetry. For the conductivity, we need to perturb the system
and investigate its linear response. The conductivity of the p-wave
superconductor is anisotropic, which makes the fluctuation analysis more
complicated than the s-wave case. We here focus on the decoupled mode
$a_y^3(t,u)$, decoupled sector $\left\{a_x^1(t,u),a_t^2(t,u),a_x^3(t,u)\right\}$
and leave the full fluctuation analysis including spatial momentum for future
investigations.

\begin{figure}[h]
\includegraphics[scale=0.8]{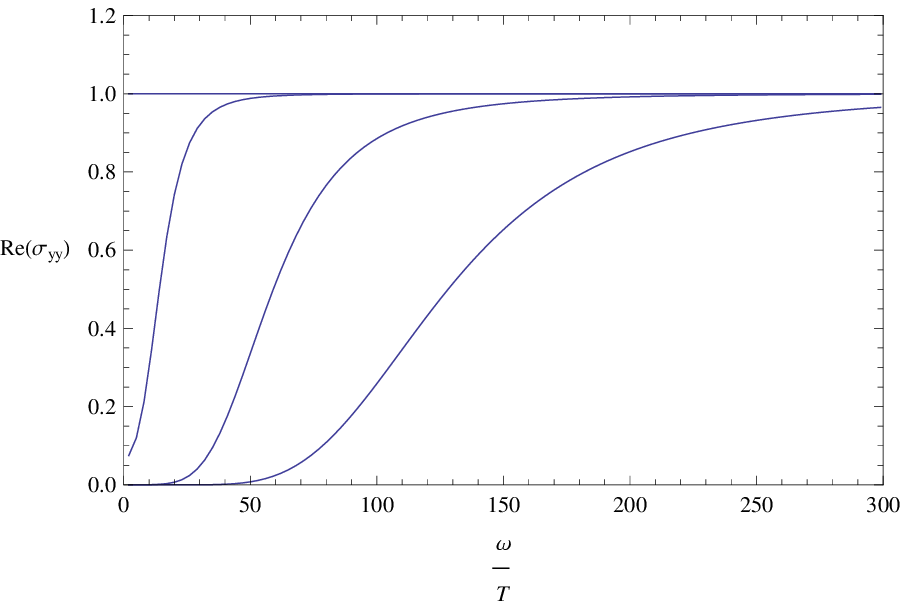}
\includegraphics[scale=0.8]{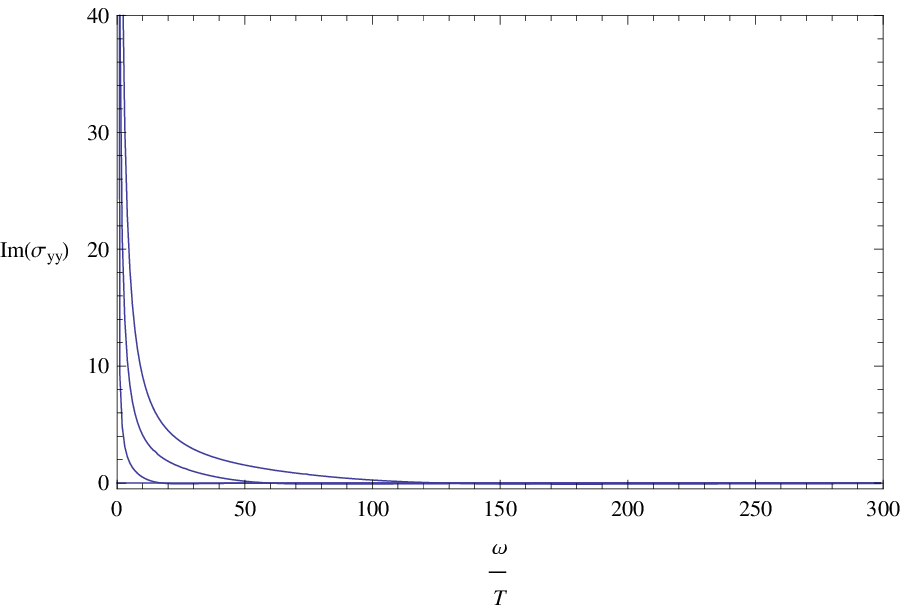}
\caption{The real (imaginary) part of the AC conductivity along the $y$
direction of the p-wave superconductor at different temperature $T/T_c=1.0,
0.588587, 0.171829, 0.079889$ from top to bottom (bottom to top).}
\label{figure10}
\end{figure}

For the fluctuation mode $a_y^3(t,u)\sim e^{-i \omega t}a_y^3(u)$, which is
decoupled from other modes, we have the following equation,
\begin{equation}
{a_y^3}^{\prime\prime}+\left[\frac{f(u)^{\prime}}{f(u)}-\frac{z-1}{u}\right]{
a_y^3}^{\prime}
+\left[\frac{u^{2z-2}\tilde{\omega}^2}{f(u)^2}-\frac{\psi^2}{f(u)}\right]
a_y^3=0.
\end{equation}
This mode looks like the electromagnetic fluctuation in the scalar system for
the s-wave model and the procedure for numerical computation of the conductivity
along the $y$ direction is of course in parallel with the s-wave situation. We
here only present the final results for the conductivity in FIG.~\ref{figure10}.

We then move on to the conductivity along the $x$ direction. For this purpose,
one should analyze the fluctuations
$\left\{a_t^1(t,u),a_t^2(t,u),a_x^3(t,u)\right\}\sim e^{-i \omega
t}\left\{a_t^1(u),a_t^2(u),a_x^3(u)\right\}$, which is also decoupled from other
modes but is self-coupled. There are three coupled equations of second order,
\begin{equation} \label{eqs axt}
\left\{ \begin{aligned}
&{a_t^1}^{\prime\prime}+\frac{z-1}{u}{a_t^1}^{\prime}+\frac{\phi\psi}{f(u)}
a_x^3=0,\\
&{a_t^2}^{\prime\prime}+\frac{z-1}{u}{a_t^2}^{\prime}-\frac{\psi}{f(u)}
\left(i\tilde{
\omega}a_x^3+\psi a_t^2\right)=0,\\
&{a_x^3}^{\prime\prime}+\left[\frac{f(u)^{\prime}}{f(u)}+\frac{1-z}{u}\right]{
a_x^3}^{\prime}-
\frac{u^{2z-2}}{f(u)^2}\left(-\tilde{\omega}^2a_x^3+\phi\psi a_t^1+i
\tilde{\omega}\psi a_t^2\right)=0,                          \end{aligned}
\right.
\end{equation}
and two constraints of first order\footnote{These constraints come form the
radial gauge we have chosen for the gauge field fluctuation, i.e., $a_u^a=0$
when deriving the equations of motion.},
\begin{equation}
\left\{ \begin{aligned}
&i\tilde{\omega}{a_t^1}^{\prime}+\phi {a_t^2}^{\prime}-\phi^{\prime}a_t^2=0,\\
&-i\tilde{\omega}{a_t^2}^{\prime}+\phi {a_t^1}^{\prime}-\phi^{\prime} a_t^1+
\frac{f(u)}{u^{2z-2}}\left(\psi {a_x^3}^{\prime}-a_x^3 \phi^{\prime}\right)=0.
\end{aligned} \right.
\end{equation}

Near the horizon $u=1$, we choose the ingoing wave boundary condition for
different modes and also impose that the time components vanish at the horizon,
\begin{equation} \label{eqs horizon}
\left\{ \begin{aligned}
&a_x^3=(1-u)^{-i\tilde{\omega}/4}\left[1+{a_x^3}^{(1)}(1-u)+{a_x^3}^{(2)}
(1-u)^2+{a_x^3}^ {(3)}(1-u)^3+\cdots\right],\\
&a_t^1=(1-u)^{-i\tilde{\omega}/4}\left[{a_t^1}^{(1)}(1-u)+{a_t^1}^{(2)}(1-u)^2+{
a_t^1}^ {(3)}(1-u)^3+\cdots\right],\\
&a_t^2=(1-u)^{-i\tilde{\omega}/4}\left[{a_t^2}^{(1)}(1-u)+{a_t^2}^{(2)}(1-u)^2+{
a_t^2}^ {(3)}(1-u)^3+\cdots\right],
\end{aligned} \right.
\end{equation}
where we have used the linearity of eqs.~(\ref{eqs axt}) to set the scale of
$a_x^3$ at the horizon to 1. The coefficients in the above expansions can be
fully determined by plugging the expansion into eqs.~(\ref{eqs axt}) and
counting powers of $(1-u)$. Then, eqs.~(\ref{eqs horizon}) can provide initial
conditions for these second order differential equations (\ref{eqs axt}). In
fact, we use these power solutions to do numerical integration from the horizon
to the conformal boundary as mentioned in the last section.

At the conformal boundary, the general solutions to eqs.~(\ref{eqs axt}) are of
the form,
\begin{equation} \label{new boundary}
\left\{ \begin{aligned}
&a_t^1={A_t^1}^{(0)}+ {A_t^1}^{(1)}\log u+\cdots,\\
&a_t^2={A_t^2}^{(0)}+ {A_t^2}^{(1)}\log u+\cdots,\\
&a_x^3={A_x^3}^{(0)}+ {A_x^3}^{(1)}u^2+\cdots
\end{aligned} \right.
\end{equation}
We can also expand the constraint equations near the conformal boundary, but it
is of no use for later calculations.

As argued in \cite{0805.2960}, for the conductivity to be a gauge invariant
quantity, we need to construct a new mode from $a_t^1, a_t^2, a_x^3$ which
should be invariant under the gauge transformation that respects our gauge
choice. We here do not go into the details of the construction of this field but
write down the mode directly,
\begin{equation}
\tilde{a_x^3}\equiv a_x^3+\psi\frac{i \tilde{\omega} a_t^2+\phi
a_t^1}{\phi^2-\tilde{\omega}^2}.
\end{equation}
Plugging the boundary behavior as in eqs.~(\ref{new boundary}) into the newly
defined mode and expanding it near the conformal boundary,
\begin{equation}
\tilde{a_x^3}=\tilde{A_x^3}^{(0)}+ \tilde{A_x^3}^{(1)}u^2+\cdots,
\end{equation}
with
\begin{eqnarray}
\tilde{A_x^3}^{(0)}={A_x^3}^{(0)},\quad
\tilde{A_x^3}^{(1)}=\frac{{A_t^1}^{(0)}}{\mu}\psi^{(2)}+{A_x^3}^{(1)}.
\end{eqnarray}
Then, the conductivity along the $x$ direction is straightforwardly defined as
\begin{equation}
\sigma_{xx}(\omega)=\frac{1}{i\omega}\frac{\tilde{A_x^3}^{(1)}}{\tilde{A_x^3}^{
(0)}}.
\end{equation}
Notice that the above formula is very different from equation (4.19) of
\cite{0805.2960} due to the appearance of the logarithmic terms in the time
component of the SU(2) gauge field (for both the background and the
fluctuations). The definition for the spectral function is similar to the s-wave
case and we list them in what follows for completeness,
\begin{equation}
\mathcal{R}=-2 \frac{\tilde{A_x^3}^{(1)}}{\tilde{A_x^3}^{(0)}}
\end{equation}
along the x-direction;
\begin{equation}
\mathcal{R}=-2 \frac{{A_y^3}^{(1)}}{{A_y^3}^{(0)}}
\end{equation}
along the y-direction.

\begin{figure}[h]
\includegraphics[scale=0.8]{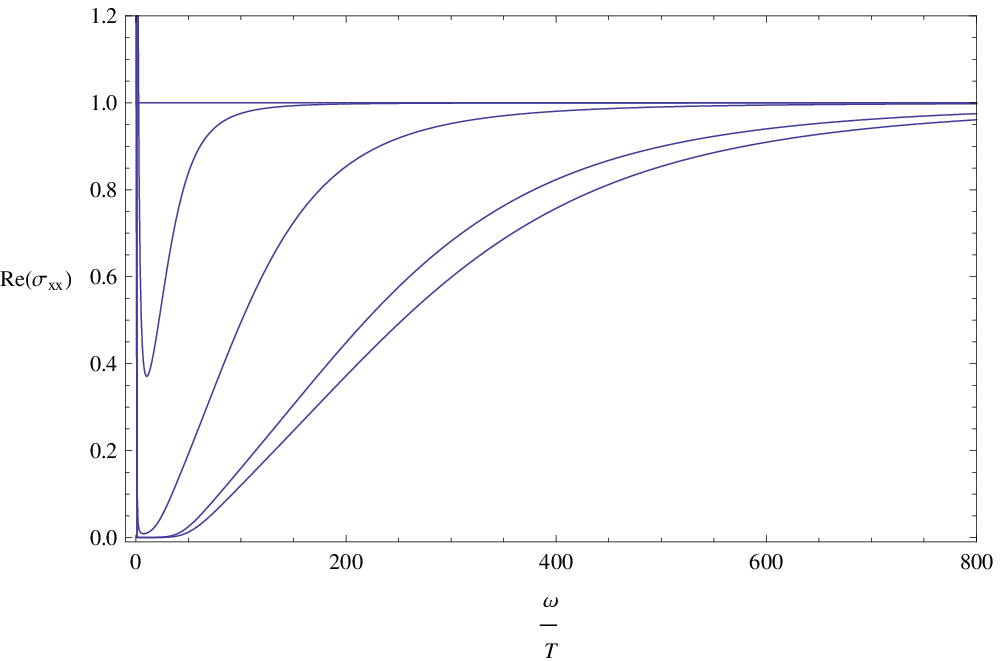}
\includegraphics[scale=0.85]{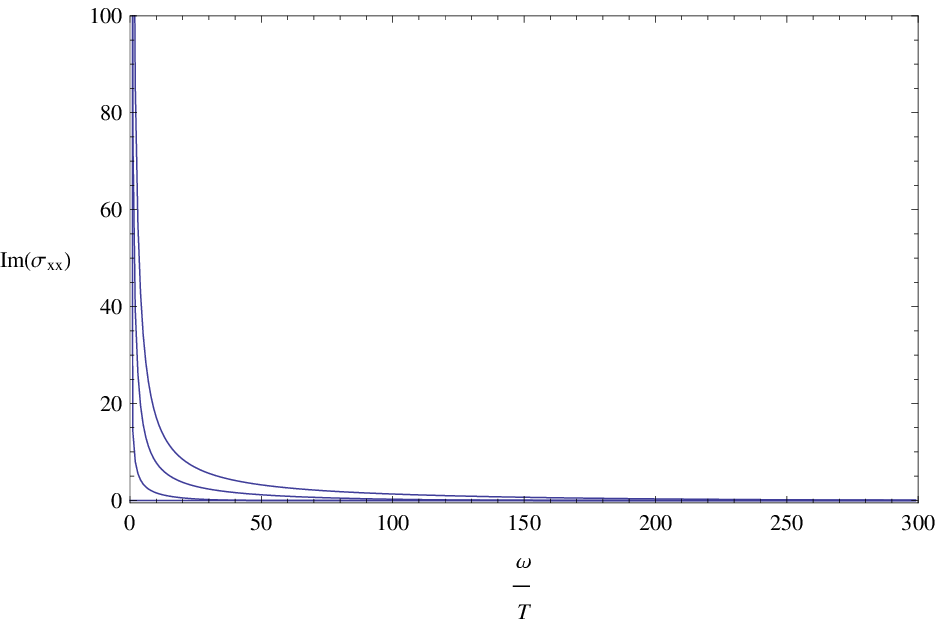}
\caption{The real (imaginary) part of the AC conductivity along $x$ direction of
the p-wave superconductor at $T/T_c=1.0, 0.588587,0.171829,0.079889,0.0691493$
from top to bottom (bottom to top).}
\label{figure11}
\end{figure}

\begin{figure}[h]
\includegraphics[scale=1.5]{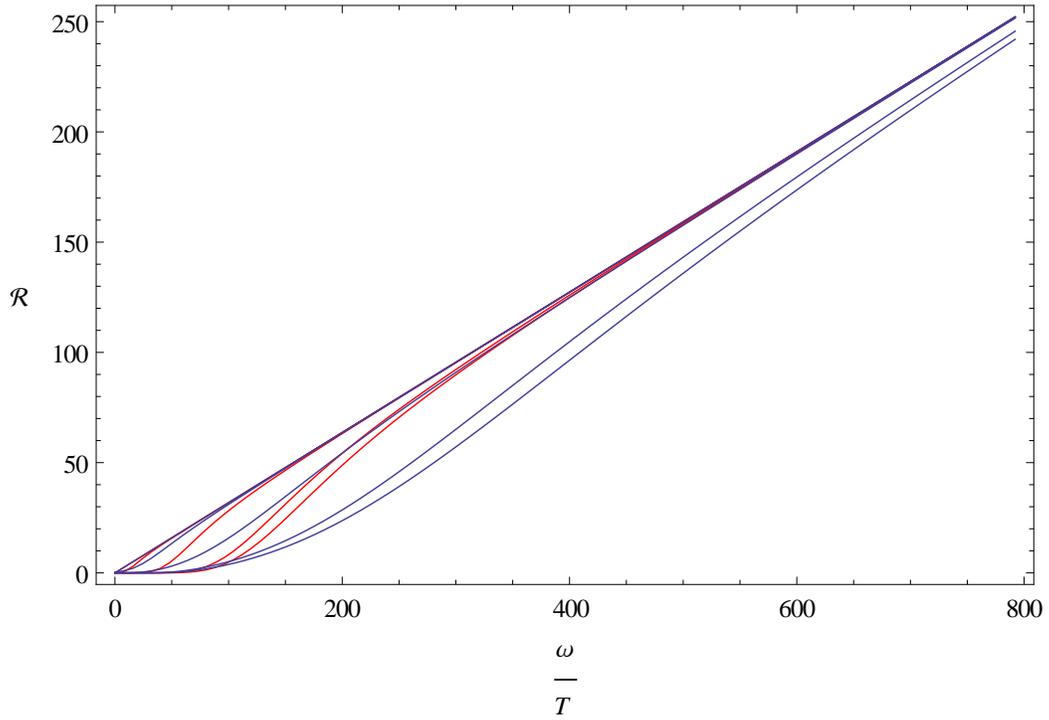}
\caption{The spectral functions for the modes $a_y^3$ (red) and $\tilde{a_x^3}$
(blue) at
temperature corresponding to the conductivity pictures.}
\label{figure12}
\end{figure}

In FIG.~\ref{figure11} we plot numerical results for the conductivity along the
$x$ direction. We can see that the results are very similar to those of the $y$
direction stated in FIG.~\ref{figure10}. Explicitly, there is a DC infinity in
the conductivity along the $x$ direction. What is strikingly different from the
$AdS_4$ p-wave superconductor is that there is no pole for the imaginary part of
the conductivity at nonzero frequency. Mathematically, this is due to the
logarithmic term appearing in the boundary behavior for the time component
fluctuations of the gauge field. We will take this as the effect of the
nontrivial dynamical exponent $z\neq1$. The real (imaginary) part of the
conductivity approaches 1 (0) more slowly than the $y$ direction case, which is
consistent with the spectral function profiles plotted in FIG.~\ref{figure12}.
This can be explained by the anisotropic characteristic of the p-wave
superconductor.

The qualitative behavior is similar to the corresponding s-wave case: at high
frequency the real part of the conductivity approaches 1 while the imaginary
part goes to 0; a peak also appears at zero frequency, indicating a DC
superconductivity; the gap forms as the temperature decreases and it gets
increasingly deep until the conductivity is exponentially small. The appearance
of the delta peak at zero frequency can also be well understood by the
Kramers-Kr\"{o}nig relation (\ref{kk relation}). Although the conductivity
$\sigma_{yy}(\omega)$ behaves qualitatively similar to $\sigma_{xx}(\omega)$,
the anisotropic characteristic of the p-wave superconductor can also be seen
from the two conductivities: the real part of the $\sigma_{xx}(\omega)$ grows
much slowly than that of $\sigma_{yy}(\omega)$. These features are common with
the AdS superconductors. One main difference is that there is no pole at nonzero
frequency for the imaginary part of $\sigma_{xx}(\omega)$ as mentioned before.
Another main difference from the AdS black hole superconductor is that the
imaginary part of the conductivity never goes below zero and approaches zero
quite slowly.
\section{Summary}\label{section5}
In this work we explored properties of holographic superconductors with
nontrivial dynamical exponent by putting the Abelian-Higgs model (s-wave) or
SU(2) gauge field (p-wave) into the Lifshitz black hole geometry constructed in
\cite{0812.0530}. We found that the order parameters $\langle\mathcal{O}\rangle$
all have mean field behavior $(T-T_c)^{1/2}$ near the critical temperature
$T_c$, which is qualitatively consistent with the AdS superconductors as well as
BCS theory. One difference between our results and previous investigations on
condensates is that the condensates do not approach some fixed values in the
zero temperature limit\footnote{We would like to state that this feature is not
due to the numerical approach used here. Actually, we already used our numerical
method to carry out similar computations in other black hole backgrounds \cite{10.1016/j.nuclphysb.2012.07.014}, of
asymptotic AdS type and found that the condensed operator approach constant
value. Therefore, we conclude that the feature found here is due to the
nontrival dynamical exponent $z$.}. We then plot the free energy difference
between the
normal and superconducting phases, which can be taken as evidence of the
occurrence of the superconducting phase transition. We also numerically compute
the AC conductivities and they nearly behave in the same way: a peak appears at
zero frequency, indicating a DC superconductivity; the gap forms as the
temperature decreases and it gets increasingly deep until the real part of the
conductivity gets exponentially small. The anisotropic characteristic of the
p-wave superconductor can be seen from the difference between
$\sigma_{xx}(\omega)$ and $\sigma_{yy}(\omega)$: their growth rates as
increasing the frequency are very different. But we do not see a pole at nonzero
frequency for the imaginary part of $\sigma_{xx}(\omega)$, which does exist in
the p-wave superconductor when taking the $AdS_4$ black hole geometry. Another
feature is that all the real parts of the conductivities approach 1 (but never
exceed this value) in contrast to those of the AdS case. With respect to this,
the imaginary parts of the conductivities approach zero in the high frequency
limit but never go below zero. We attribute these differences from the AdS case
to the effect of nontrivial dynamical exponent. More specifically, the black
hole geometry considered in this work is anisotropic between space and time,
very different from the Schwarzschild-AdS black hole, which results in different
asymptotic behaviors of temporal and spatial components of gauge fields than
previous conclusions in Schwarzschild-AdS black hole. These common features also
imply that general gauge/gravity duality is a useful and powerful tool in
producing some universal properties of strongly coupled system in condensed
matter physics.
\begin{acknowledgments}
The author would like to thank Johanna Erdmenger, Jonathan Shock, Xin Gao and
Da-Wei Pang, Xu Zhang for useful discussions. This work was supported by the MPS-CAS
Doctoral Training program.
\end{acknowledgments}

\end{document}